\documentclass[]{aa}  

\usepackage{color}
\usepackage{natbib}
\usepackage[english]{babel}

\usepackage{xspace}
\usepackage{siunitx}
\usepackage{txfonts}

\newcommand{\gtapprox}{\raisebox{-0.5ex}{$\,\stackrel{>}{\scriptstyle\sim}\,$}}

\newcommand\Lya{${\rm Ly\alpha}$\xspace}

\newcommand\Muv{M$_{\rm 1500}$\xspace}
\newcommand\flya{$f_{\rm Ly\alpha}$\xspace}
\newcommand\fuv{$f_{\rm uv}$\xspace}
\newcommand\sfrlya{SFR$_{\rm Ly\alpha}$\xspace}
\newcommand\sfruv{SFR$_{\rm uv}$\xspace}
\newcommand{\rf}{$f_{\rm Ly_{\alpha}}/f_{\rm UV}$\xspace}
\newcommand{\ewlya}{EW$_{\rm Ly\alpha}$\xspace}

\begin{document} 

\title{MUSE observations towards the lensing cluster A2744: Intersection between the LBG and LAE populations at z$ \sim $3-7}
\titlerunning{Intersection between the LBG and LAE population at z $\sim$ 3-7 behind lensing field A2744}


 \author{G.~de La Vieuville\inst{\ref{inst1}}
   \and
R.~Pell\'o\inst{\ref{inst11},\ref{inst1}} \and
J.~Richard\inst{\ref{inst2}}  \and
G.~Mahler\inst{\ref{inst3}}  \and
L.~L\'ev\^eque\inst{\ref{inst1}} \and
F. E.~Bauer\inst{\ref{inst5}}\inst{\ref{inst6}}\inst{\ref{inst7}} \and
D. J.~Lagattuta\inst{\ref{inst2}}  \and
J.~Blaizot\inst{\ref{inst2}} \and
T.~Contini\inst{\ref{inst1}} \and
L.~Guaita\inst{\ref{inst5}} \and
H. Kusakabe\inst{\ref{inst8}} \and
N.~Laporte\inst{\ref{inst1},\ref{inst12},\ref{inst13}} \and
J.~Martinez\inst{\ref{inst2}}  \and
M. V.~Maseda\inst{\ref{inst10}} \and
D.~Schaerer\inst{\ref{inst8}} \and
K. B.~Schmidt\inst{\ref{inst9}} \and
A.~Verhamme\inst{\ref{inst8}}}


\institute{
Institut de Recherche en Astrophysique et Planétologie (IRAP), Université de Toulouse, CNRS, UPS, CNES, 14 Av. Edouard Belin, F-31400 Toulouse, France,  \email{geoffroy.dlv@gmail.com, roser.pello@lam.fr}\label{inst1}
 \and
 Aix Marseille Université, CNRS, CNES, LAM (Laboratoire d’Astrophysique de Marseille), UMR 7326, 13388, Marseille, France\label{inst11}
 \and
Univ Lyon, Univ Lyon1, Ens de Lyon, CNRS, Centre de Recherche Astrophysique de Lyon (CRAL) UMR5574, F-69230, Saint-
Genis-Laval, France\label{inst2}
\and
Department of Astronomy, University of Michigan, 1085 S. University Ave., Ann Arbor, MI 48109, USA\label{inst3}
Department of Physics and Astronomy, University College London, Gower Street, London WC1E 6BT, UK\label{inst4}
\and
Instituto de Astrof\'isica and Centro de Astroingenier\'ia, Facultad de F\'isica, Pontificia Universidad Cat\'olica de Chile, Casilla 306, Santiago 22, Chile\label{inst5}
\and
Space Science Institute, 4750 Walnut Street, Suite 205, Boulder, Colorado 80301, USA\label{inst6}
\and
Millenium Institute of Astrophysics (MAS), Nuncio Monse{\~{n}}or S{\'{o}}tero Sanz 100, Providencia, Santiago, Chile\label{inst7}
\and
Observatoire de Gen\'eve, Université de Genève, 51 Ch. des Maillettes, 1290 Versoix, Switzerland\label{inst8}
\and
Cavendish Laboratory, University of Cambridge, 19 JJ Thomson Avenue, Cambridge CB3 0HE, UK\label{inst12}
\and
Kavli Institute for Cosmology, University of Cambridge, Madingley Road, Cambridge CB3 0HA, UK\label{inst13}
\and
Leiden Observatory, Leiden University, P.O. Box 9513, 2300 RA, Leiden, The Netherlands\label{inst10}
\and
Leibniz-Institut für Astrophysik Potsdam (AIP), An der Sternwarte 16, D-14482 Potsdam, Germany\label{inst9}
}

   \date{Received 3rd February 2020 ; accepted  21st September 2020}

 
   \abstract{We present a study of the intersection between the populations of star forming galaxies (SFGs) selected as either Lyman break galaxies (LBGs) or Lyman-alpha emitters (LAEs) in the redshift range $2.9 - 6.7$, within the same volume of universe sampled by the Multi-Unit Spectroscopic Explorer (MUSE) behind the Hubble Frontier Fields lensing cluster A2744. We define three samples of star-forming galaxies: LBG galaxies with an LAE counterpart (92 galaxies), LBG galaxies without LAE counterpart (408 galaxies) and LAE galaxies without an LBG counterpart (46 galaxies). All these galaxies are intrinsically faint due to the lensing nature of the sample (\Muv$\ge$-20.5). The fraction of LAEs among all selected star-forming galaxies increases with redshift up to $z \sim 6$ and decreases for higher redshifts, in agreement with previous findings. The evolution of  LAE/LBG populations with UV magnitude and \Lya luminosity shows that the LAE selection is able to identify intrinsically UV faint galaxies with \Muv$\ge$-15 that are typically missed in the deepest lensing photometric surveys. The LBG population seems to fairly represent the total population of star-forming galaxies down to \Muv$\sim-15$. Galaxies with \Muv$<-17$ tend to have \sfrlya$<$\sfruv, whereas the opposite trend is observed within our sample for faint galaxies with \Muv$>-17$, including galaxies only detected by their \Lya emission, with a large scatter. These trends, previously observed in other samples of star-forming galaxies at high-$z$, 
  are seen here for very faint \Muv$\sim -15$ galaxies, much fainter than in previous studies.

There is no clear evidence, based on the present results, for an intrinsic difference on the properties of the two populations selected as LBG and/or LAE. The observed trends could be explained by a combination of several facts, like the existence of different star-formation regimes, the dust content, the relative distribution and morphology of dust and stars, or the stellar populations.}


   \maketitle
%
   

\section{Introduction}

   Intrinsically faint star-forming galaxies (hereafter SFGs) are presently considered as the main sources responsible for cosmic reionization. Ultra-deep photometry obtained by the {\it Hubble Space Telescope} (\textit{HST}) on blank fields combined with ground-based photometry and spectroscopy has fundamentally improved our knowledge of the galaxy UV Luminosity Function (LF) up to $z \sim $ 10
\citep[see e.g.][and the references therein]{Bouwens2004,McLure2013,Bouwens2015b,
     Finkelstein2015}.
The integration of the UV LF is used to derive the evolution of the cosmic star-formation density, and the density of ionizing radiation (usually assuming a constant star-formation rate), while the two key parameters being the slope of the faint-end of the LF and the faint-end integration limit \citep[see e.g.][]{Bouwens2015a}.
Up to $ z \sim$7, current observations reach as deep as \Muv$\sim-17$ in blank fields, that is about three magnitudes brighter than the faint-end UV luminosity limit which is needed to reionize the universe at $z \sim$ 6-7. Using lensing clusters as gravitational telescopes makes it possible to reach \Muv$\sim-15$ at $z \sim$7 therefore improving our constraints on the contribution of SFGs to reionization (See for example the work done in \cite{Atek2015} and \cite{Livermore2017} using data from the {\it Hubble Frontier Fields} (HFF) project \citep{Lotz2017}).

However, the ability of SFG to reionize the universe depends not only on the faint-end slope of the UV LF (and its actual shape) and the faint-end integration limit, but also on the escape fraction of ionizing radiation. In addition, most samples used for this exercise, either in blank fields or in lensing fields, are photometrically selected and have few if any spectroscopic redshifts available. The interrelation between the different SFG populations, selected as Lyman-Break Galaxies (LBGs) or Lyman Alpha Emitters (LAEs), has been a hot topic for many years, both from the observational and the theoretical points of view
\citep[see e.g.][]{Nagamine2010,Pentericci2011,Shapley2011,Stark2011,Dayal2012,Erb2014,Orsi2014,Tilvi2014,Schenker2014,Garel2016,deBarros2017,Caruana2018,ArrabalHaro2018}. 

The same holds for the relative escape fractions of \Lya and UV photons  (i.e., the ability of observed quantities such as \Lya and rest-frame UV fluxes to trace the ionizing radiation and noted hereafter \flya and \fuv) \citep[see e.g.][and the references therein]{Hayes2015,
Steidel2018}.

Nevertheless, observations of LBG and LAE are rarely carried out in the same volume of universe due to observational limitations.
Recent pioneering studies have started to address this issue in blank fields, such as \cite{ArrabalHaro2018} 
using the SHARDS Survey of the GOODS-N field, based on deep imaging survey using 25 medium band filters, or \citet{Inami2017}
\cite{Hashimoto2017}, \cite{Maseda2018} and \cite{Kusakabe2020}, 
all based 
on ultra-deep IFU data on the {\it Hubble Ultra Deep Field} (HUDF) from MUSE
\citep[the Multi-Unit Spectroscopic Explorer][]{Bacon2010,Bacon2017}.

   In this paper, we investigate the intersection between the populations of SFGs selected as LBGs and LAEs in the range $2.9 < z < 6.7$, within the same volume of universe sampled by MUSE behind the HFF lensing cluster Abell 2744 (hereafter A2744).
For the first time, such a combined survey is performed behind a lensing cluster, using the deepest images available from the HFF. Taking advantage of the magnification provided by the lensing cluster, the survey reaches an average depth of \Muv$\sim -15$ and probes galaxies as faint as \Muv $\sim -12$ in some areas, and \Lya luminosities in the range $40 \lesssim \log(L_{Ly\alpha}) \lesssim 43$ . We investigate the prevalence of \Lya emission among the faintest LBG population (\Muv $ >-20.25$), as well as the relationship between \Lya and UV luminosity. 
Hereafter the term LBG is used to mean galaxies identified by their UV continuum, using whether the classical color-color diagrams \citep[see e.g.][] {Steidel_2003, 
2007ApJ...671.1212O, 
Bouwens2015b} 
or photometric redshifts  \citep[see e.g.][] {McLure2009, 
2018A&A...620A..51P}. 
In Sect.\ \ref{observations} we briefly describe the relevant MUSE and HFF data available for this study. Sect.\ \ref{selection} presents the selection of LBG and LAE samples at $z\sim$ 3-7 behind A2744. The results on the intersection of LBG and LAE populations are presented and discussed in Sect.\ \ref{results}. Conclusions are given in Sect.\ \ref{conclusion}. 

The cosmology adopted throughout this paper is  
$\Omega_{\Lambda}=0.7$, $\Omega_{m}=0.3$  and $H_{0}=$ 70\ km\ s$^{-1}$\ Mpc$^{-1}$
All magnitudes are given in the AB system \citep{Oke1983}.

\section{MUSE and HFF data of A2744}
\label{observations}

MUSE integral-field observations of A2744 were carried out as part of the MUSE  Guaranteed Time Observing (GTO) program on lensing clusters (GTO Program 094.A-0115; PI: Richard)
\citep[see][for additional information.]{Mahler2018,deLaVieuville2019}
The field observed was a 2$\times$2 MUSE mosaic covering the entire multiple-image area, with exposure times ranging between 3.5 and 5 hours per pointing, plus a central pointing with 2 additional hours of exposure centered at $\alpha$=00:14:20.95 $\delta$=-30:23:53.9 (J2000). All details regarding MUSE data reduction, source detection process and mass model construction can be found in \cite{Mahler2018}. Throughout this study we use the gold mass model presented in Table 4 of \citet{Mahler2018} which has an average RMS of multiple images in the image plane of 0.67\arcsec. The MUSE catalogue includes 171 LAEs \footnote{publicly available at \url{http://muse-vlt.eu/science/a2744/}} before multiple image removal, with redshifts in the range $2.9 < z < 6.7$. All LAEs were detected with \textsc{Muselet}\footnote{\textsc{Muselet} is part of the python MPDAF package \citep{Piqueras2017}}, a detection software for emission lines in MUSE cubes. Table~\ref{tab_fraction_LAE} summarizes the effective lens-corrected volume surveyed by the MUSE observations behind A2744. The detection fluxes of the LAEs are measured with \textsc{SExtractor} \citep{Bertin1996} using the \verb+MAG_AUTO+ segmentation on \Lya Narrow Band (NB) images. These NB images are continuum subtracted, and their spectral width is manually adjusted to the spectral extent of each LAE to ensure the best recovery of the total \Lya flux. The automated segmentation is efficient in dealing with a wide variety of spatial profiles, including high distortion induced by lensing. All technical details regarding the LAE extraction method including some examples can be found in Sect. 3.2 of \citet{deLaVieuville2019}.

The selection of LBGs in this field is based on the HFF observations of A2744 (ID: 13495, P.I: J. Lotz). Seven filters are available for Hubble Space Telescope data, three from the Advanced  Camera  for  Surveys (ACS: F435W, F606W, F814W), and four from the Wide Field Camera 3 (WFC3: F105W, F125W, F140W, and F160W). In this study we use self-calibrated  data  provided  by  STScI (version  v1.0 for WFC3 and v1.0-epoch2 for ACS), with a pixel size of 60 mas. The full MUSE mosaic is contained within all these seven HFF bands. We used the photometric catalog released by Astrodeep \citep{Merlin2016,Castellano2016}, which also includes imaging from VLT/Hawk-I K-band and the first two \textit{Spitzer}/IRAC bands. For each of the seven \textit{HST} filters, both the intra-cluster light (ICL) and bright-cluster galaxies (BCGs) were modeled and subtracted. The photometry was measured on these processed images with SExtractor \citep{Bertin1996}, with PSF matching techniques using high spatial resolution images as priors for the source segmentation.
The complete procedure is detailed in \cite{Merlin2016}.
The complete catalogue has 3587 entries for the entire A2744 lensing field, of
which 2596 are detected in the F160W image, a further 976 are detected in a weighted stack of F105W, F125W, F140W and F160W and undetected in F160W alone, and 15 are BCGs. Note that the Astrodeep catalog is NIR-selected by construction, and this fact could have some implications on the subsequent results.

   Before any comparison with the MUSE source catalogue, the entire Astrodeep catalogue was filtered to remove untrustworthy photometry points and/or sources. Since VLT/Hawk-I, IRAC 1 and IRAC 2 have larger PSFs than the \textit{HST} filters, the photometry computed in these three bands is more often contaminated
by nearby galaxies. Following the flags given in the catalogue, all photometric entries likely to be affected by this effect (indicated by flag COVMAX; see \cite{Merlin2016}) were given an upper detection limit in these filters when computing photometric redshifts. We also removed 220 sources flagged as likely residuals of the foreground subtraction (SEXFLAG $>$ 16 and VISFLAG $>$ 1). Finally, the catalogue was cut to the exact MUSE FoV, leading to a final selection of 2666 sources detected in the Astrodeep catalogue within the MUSE mosaic.

   In order to build a single coherent catalogue, a cross match was performed between the remaining 2666 Astrodeep sources and the 171 LAEs of the MUSE catalogue, using a matching radius of
$r = 0.8\arcsec$ (4 MUSE pixels). In case several entries are pointing towards the same source, only the closest match is kept. The difference in morphology between the UV and \Lya emission \citep[see e.g.][]{Leclercq2017,Wisotzki2018} and the distortions induced by lensing are likely to be the dominant causes of mismatch. For this reason, all LAEs and their closest Astrodeep counterpart were manually inspected to confirm or reject the match. In the case of multiple-image systems, it is always possible to select the less ambiguous image of the system when assembling the final sample (see below).  
At the end of the matching process, the merged catalogue contains 2724 galaxies of which 113 are seen in both the MUSE and Astrodeep catalogues, and 58 are LAEs with no detected UV counterpart in the Astrodeep catalogue.

\section{Selection of LBG and LAE galaxies at z$\sim$2.9-6.7}
\label{selection}

Two different methods were used to select LBGs in this field among the sources detected by Astrodeep in the MUSE field of view: the usual three-band dropout technique applied to HFF data, and a method based on pure photometric redshifts and probability distributions. For the dropout (two-color) technique, we adopted the same criteria as proposed by \citet{Bouwens2015b} for the selection of galaxies at $z \sim$3.8, 4.9, 5.9, and 6.8. Photometric redshifts and redshift probability distribution (noted $P(z)$) were computed with the code
{\it New$-$Hyperz\/}\footnote{http://userpages.irap.omp.eu/$\sim$rpello/newhyperz/}, 
originally developed by \citet{Bolzonella2000}, based on the fitting of
the photometric Spectral Energy Distributions (SED) of galaxies.
The input Astrodeep catalog presented in Sect.\ \ref{observations} also includes photometric redshifts. However, a new evaluation of the photometric redshifts is carried out here in order to preserve the consistency of the fit obtained for the different parameters used in this study.

\begin{figure}
   \centering
   \includegraphics[width=\hsize]{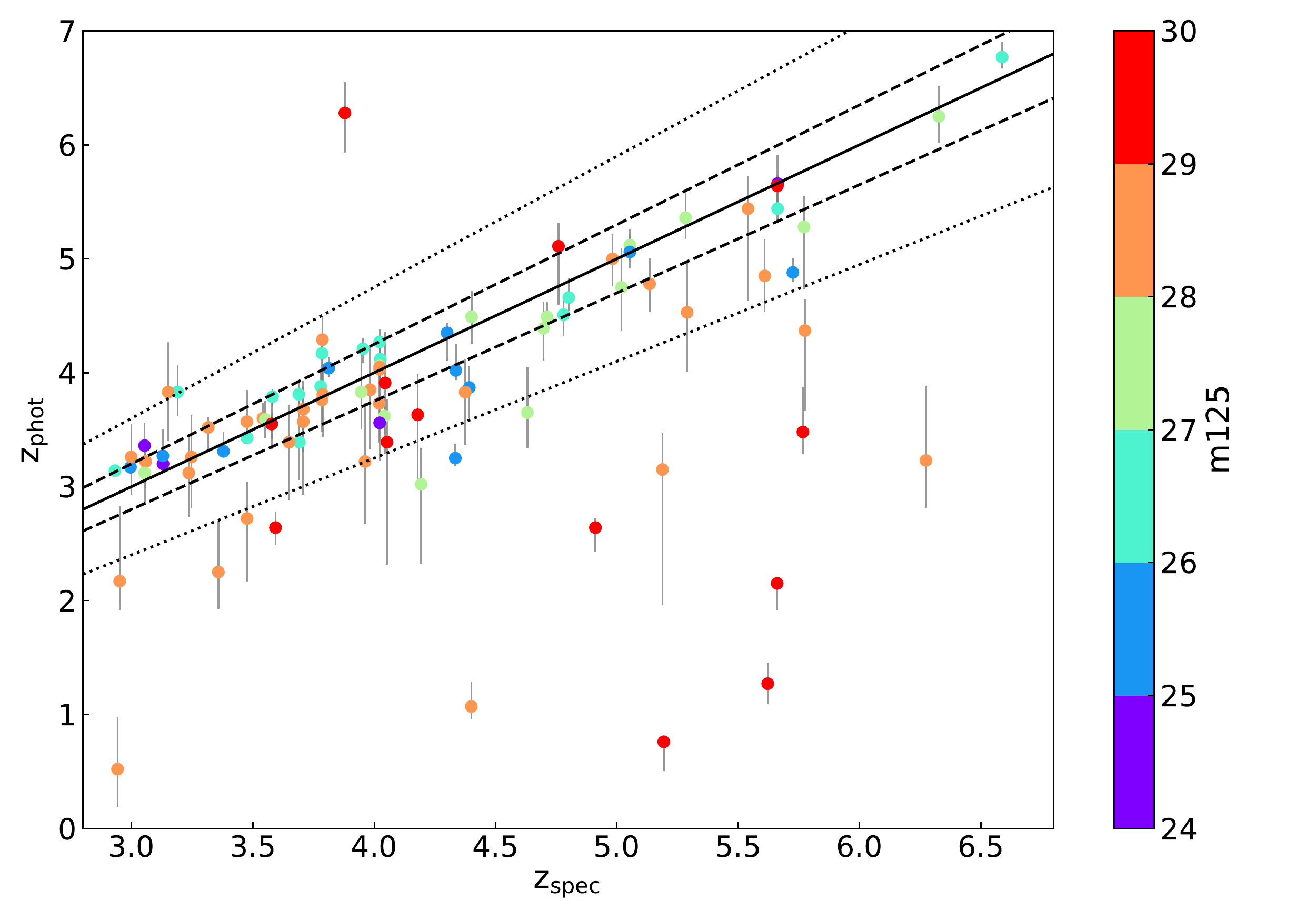}
   \caption{Comparison between photometric and spectroscopic redshifts for galaxies spectroscopically confirmed at $ z \sim$3-7. Long-dashed and dashed lines display the locus of $z_{\rm phot}=z_{\rm spec}\pm 0.05(1+z)$ and $\pm 0.15(1+z)$ respectively, and solid black line is the one-to-one relation. Colors encode the observed magnitudes in the F125W filter, showing that a vast majority of galaxies with poor photometric redshifts have $m($F125W$)\ge 28$.}
         \label{fig_zz}
   \end{figure}

Photometric redshifts with {\it New$-$Hyperz\/} were computed in the range $z =$[0,8]. The template library used in this study includes 14 templates: 
eight evolutionary synthetic SEDs from the Bruzual \& Charlot code \citep{Bruzual2003}, 
with Chabrier IMF \citep{Chabrier2003} and solar metallicity;
a set of four empirical SEDs compiled by \citet{Coleman1980};
and two starburst galaxies from the \cite{Kinney1996} library.
Internal extinction is considered as a free parameter following the
\cite{Calzetti2000} extinction law, with A$_V$
ranging between 0 and 1.0 magnitudes, and no prior in luminosity.
At this stage, galaxies with photometric redshifts higher than $z \ge$2.9 and integrated probability distributions $P(z>2.9)>60\%$ were selected as LBGs. In all cases, a S/N higher than 3$\sigma$ was required at this stage in at least one of the filters encompassing the rest-frame UV. The quality of photometric redshifts is assessed by directly comparing the results for the galaxies with known spectroscopic redshifts in the LBG sample. 
Outliers are defined as sources with $|z_{\rm spec}-z_{\rm phot}| >
0.15(1+z_{\rm spec})$. The average accuracy reached excluding outliers is $\Delta z/(1+z)  = - 0.011 \pm 0.053$, with a median at $\Delta z/(1+z)=0.001$.
As seen in Fig.~\ref{fig_zz}, the vast majority of galaxies with poor photometric redshifts have $m(\text{F125W})\ge 28$ (or a S/N $<$5 in this filter). It is also worth noticing that good photometric redshifts could be obtained for galaxies with $z \gtapprox 2.9$, that is fully covering the redshift domain where the \Lya line can be detected by MUSE.

Given the trends found for the quality of photometric redshifts computed with {\it New$-$Hyperz\/}, a S/N higher than 5 was requested in at least one of the seven \textit{HST} filters for sources without spectroscopic redshifts. For galaxies with a spectroscopic confirmation, no S/N criterion was imposed. 
This means that nine among the 92 galaxies spectroscopically selected have a S/N lower than 5, typically ranging between 3 and 5 in several filters, a percentage small enough to preserve the results of the subsequent analysis (see Sect. \ref{results}).
Using this blind photometric redshift procedure, and the S/N criterion, 536 galaxies are selected as LBGs. 

For comparison, the three-band dropout technique selects 383 objects, all of them included within the sample obtained with blind photometric redshifts, with only ten exceptions. Inspection of these ten objects revealed that they have either unreliable photometric points, resulting in poorly constrained or undefined photometric redshift, or that they have $z_{\rm phot}\sim 2.9$ with a large part of their $P(z)$ under the $z =2.9$ threshold, therefore failing the $P(z>2.9) \ge 60\%$ criterion.

Because we want to be as inclusive as possible in our selection, all galaxies selected from their UV continuum and photometric redshift as described above are considered in the rest of this study. For the sake of simplicity, we will continue using the term "LBG'' to refer to this photo-z sample. And even though the presence of a break is not strictly required in our selection, it remains the main feature picked up by {\it New$-$Hyperz\/} to compute the photometric redshift for most of our galaxies.

Finally, all images were inspected to identify multiple systems. For LAEs, a robust identification of multiple systems is already provided in the MUSE catalogue. For LBGs with no LAE counterparts, the identification of multiple system is done using \textsc{Lenstool} \citep{Kneib1996,Jullo2007,Jullo2009} and the predictions of the mass models presented in \cite{Mahler2018}. Therefore, for all identified multiple image systems (both LAEs and LBGs) we only keep one source counterpart, to avoid including several times the same object in our analysis. In case one LAE image of a system matches with an UV counterpart and the other(s) do not, the one with the UV counterpart is kept.

\section{The intersection between LBG and LAE populations}
\label{results}

\begin{figure*}
   \centering
   \includegraphics[width=0.90\textwidth]{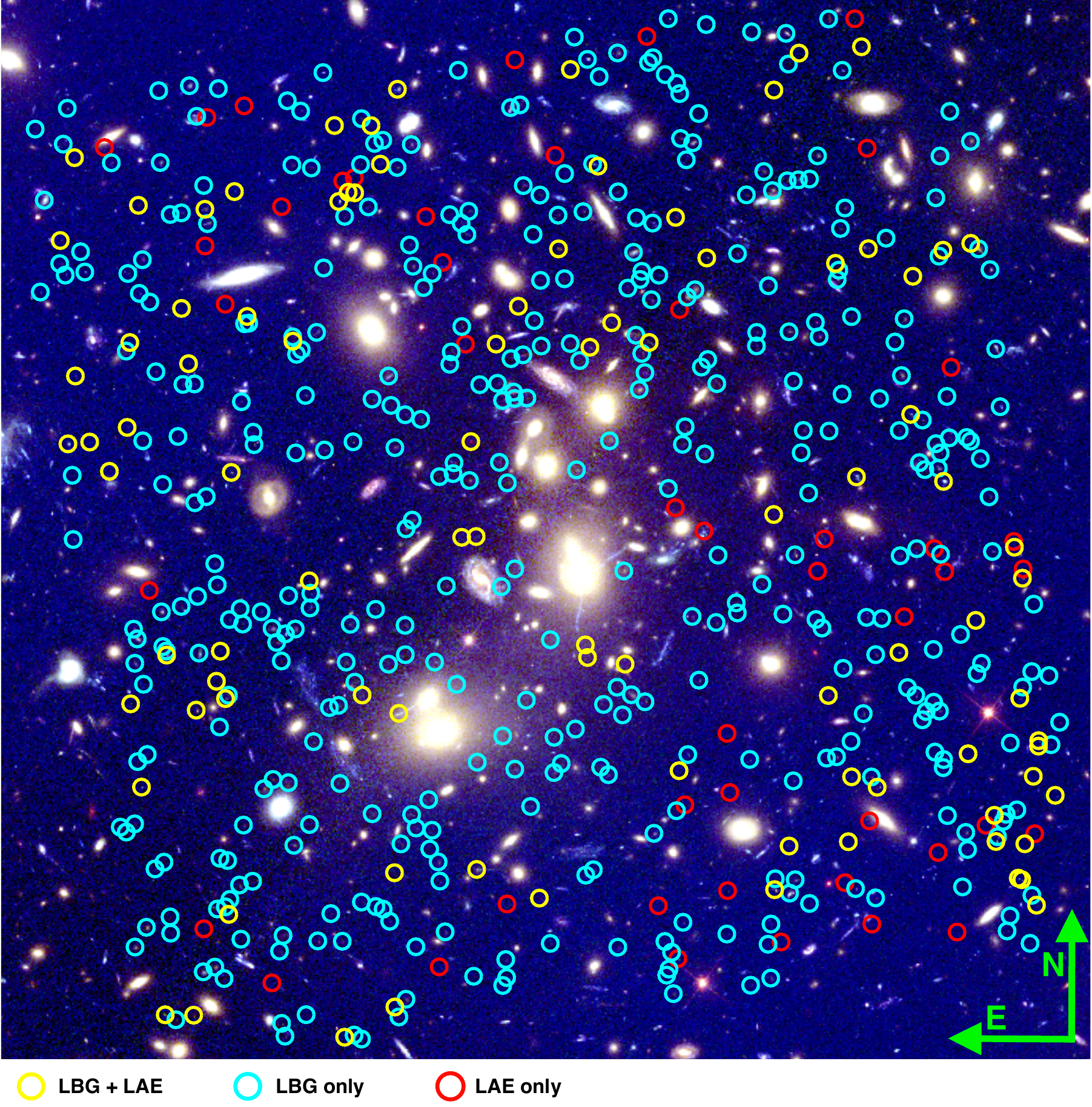}
   \caption{False-color image of A2744 showing the area covered by the MUSE observations, obtained by combining the following HFF filters: F435W (blue), F606W (green) and F814W (red). The different populations selected in this field at $2.9 < z < 6.7$ are displayed as follows: LBG without LAE counterpart in cyan (408), LBG with LAE counterpart in yellow (92) and LAE without LBG counterpart in red (58). Circles are 1.5\arcsec in diameter.}
         \label{fig_field}
   \end{figure*}

The following samples are defined in the MUSE field behind A2744 after the selection process and multiple image removal presented in Sect. \ref{selection}:
\begin{itemize}
 \item Sample 1: LBG galaxies with \Lya emission (hereafter LBG + LAE). 92 galaxies are included in this sample.   
 \item Sample 2: LBG galaxies without \Lya emission, including 408 galaxies selected by their UV rest-frame continuum.
 \item Sample 3: LAE galaxies without an LBG counterpart. This sample includes 46 galaxies. 
\end{itemize}
Fig.~\ref{fig_field} displays the different populations selected in this field overlaid on a false color image of A2744. 
In practice, the matching between the centroids of LBG and LAE galaxies is usually better than 0.1\arcsec (average value in the LBG + LAE sample). In the selection above, one concern is that some galaxies would be seen as pure LAEs because their \Lya emission can be detected by MUSE overlaid on some bright foreground galaxy, whereas their UV continuum cannot be detected. Such identification of pure LAEs would be completely artificial as it would not be representative of the intrinsic properties of the source but only of the foreground contamination. The risk has been limited by the selection of the most representative image in multiple-imaged systems (when possible), as explained in Sect.\ \ref{selection}. Indeed, as
displayed in Fig.~\ref{fig_field}, none of the pure LAEs (red circles) falls on top of a bright foreground galaxy that would prevent the detection of an underlying UV continuum.

   In order to determine the values of the UV continuum, in particular the continuum emission at the wavelength of the \Lya line (hereafter referred as ``\Lya continuum'') and the absolute magnitudes \Muv, a SED-fitting process was adopted using star-forming and young star-bursts templates (age $<$1Gyr) extracted form the Starbursts99 library \citep{Leitherer1999}. To estimate the uncertainty on both the \Lya continuum and \Muv, we performed  Monte Carlo (MC) iterations by drawing the photometry points within their error bars and iteration of the SED-fitting process.
In practice, for sources without spectroscopic redshifts, {\it New$-$Hyperz} was run two times: the first time to determine the best photometric solution and $P(z)$ as described in Sect.\ \ref{selection}, and a second time using the results of the first run to fix the best-fit redshift and perform the MC iterations. When comparing the blind photo-z obtained during the first pass to the one derived with the Starbursts99 library of templates (leaving the full redshift range free), they are found to be in full agreement for all galaxies selected as LBGs. For sources with spectroscopic redshifts, the redshift was fixed to the spectroscopic value during MC iterations. 

\begin{figure}
   \centering
   \includegraphics[width=\hsize]{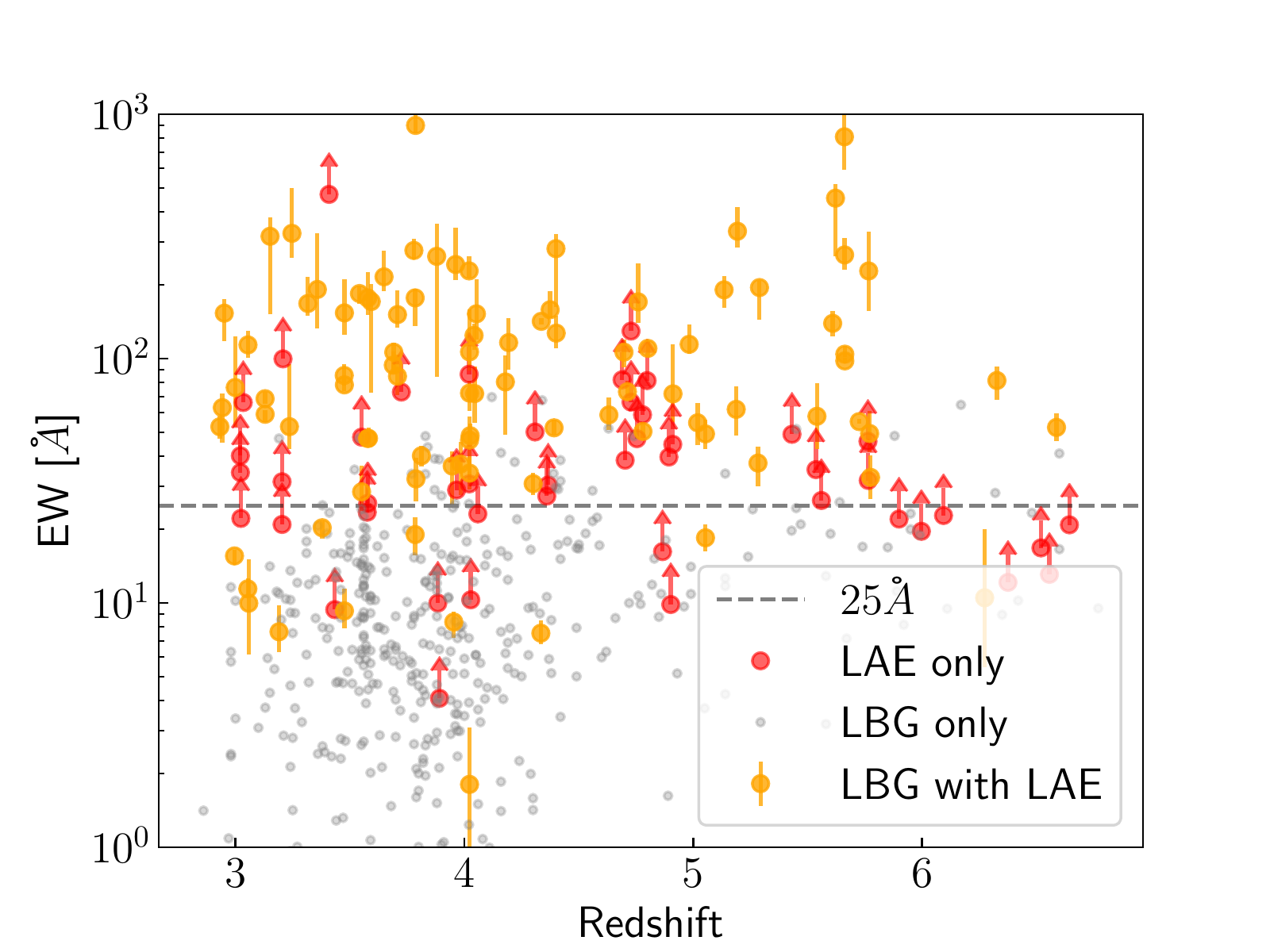}
   \caption{Redshift distribution of the rest-frame EW for \Lya emission in the three samples. For the LBG without LAE counterpart in grey, the values are upper limits and the error bars have been omitted for clarity. The dashed horizontal line corresponds to \ewlya =25  \AA.}
         \label{fig_EW}
\end{figure}

For the galaxies selected as both LAE and LBG, it is possible to compute their \ewlya using their detection flux and the \Lya continuum measured from the best-fit SEDs. For galaxies selected as pure LAEs and pure LBGs, only lower and upper limits of the \ewlya can be determined respectively. The distribution of the rest frame \ewlya with redshift for the three samples is presented in Fig.~\ref{fig_EW}. Since the magnification affects both the continuum and the \Lya emission in the same way, no additional correction is needed. For the LBG with no LAE counterpart (Sample 2), the upper limit of their \Lya emission is assumed to be a constant flux of F(Ly$\alpha$) = 1$\times$10$^{-18}$ erg/s/cm$^{2}$.
In the following, the reader should keep in mind that this is a somewhat arbitrary estimate corresponding to the flux limit above which more than 90$\%$ of the LAEs detected behind A2744 are found. A flux value ranging between 2 and 5$\times$10$^{-18}$ erg/s/cm$^{2}$, depending on the spatial and spectral profile of the source, could be used to secure a detection limit which is close to 100$\%$ completeness
\citep[see Fig. 9 in ][]{deLaVieuville2019}.  
For galaxies selected as pure LAEs in Sample 3, a limiting flux for the \Lya continuum and the corresponding apparent magnitude $m_{1500}$ were obtained for each object based on the noise of the sky background measured on the top of the area where the \Lya emission was detected, using the photometric images encompassing these wavelengths. As Sample 3 LAEs are relatively compact, a compact PSF-dominated morphology was also assumed for these galaxies in the continuum, with a maximum of 2$\sigma$ above the local sky background. This procedure allows us to take into account the different detection conditions for the continuum affecting these galaxies, in particular close to the cluster core.

\subsection{Evolution with redshift}
\label{evol_with_redshift}

\begin{figure}
   \centering
   \includegraphics[width=\hsize]{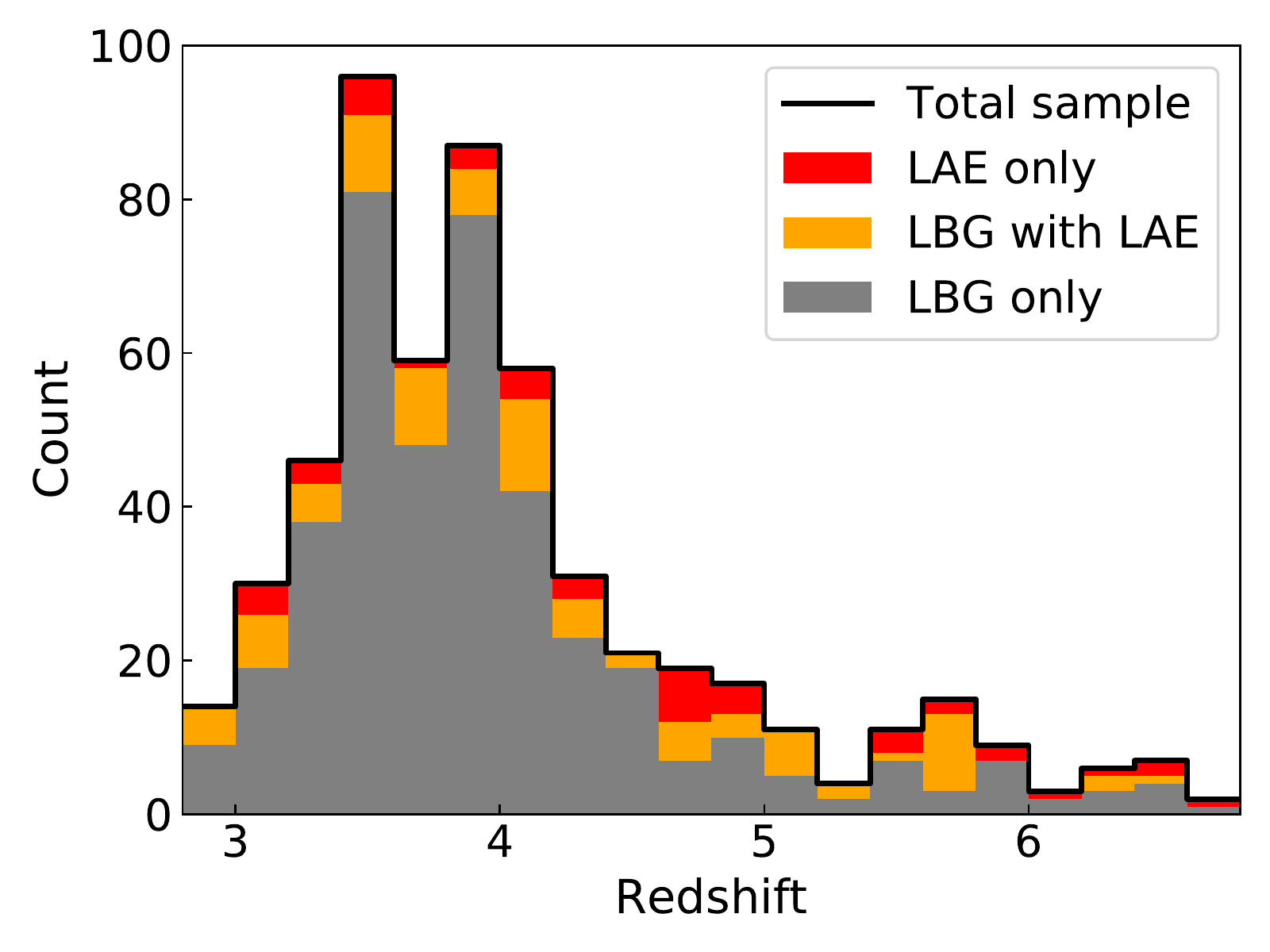}
   \caption{Redshift distribution of the three samples defined in this study.
 See text in Sect. \ref{selection} for more details on the selection of the three samples.}
         \label{fig_histoz}
   \end{figure}

Redshift histograms for the three samples considered in this study are shown stacked on top of each other in Fig.~\ref{fig_histoz}. Table~\ref{tab_fraction_LAE} displays the same information but with the redshift bins used in \cite{deLaVieuville2019} for the computation of the LAE LF. As seen in the Table and figure, the proportion of LAEs (i.e. LBG + LAE and LAE only) among SFGs increases with redshift despite the fact that these are the deepest images available in the HFF survey. 
We acknowledge that these values depend on the relative depth of the photometric and spectroscopic surveys, and as such they only indicate an observed trend for our data set.

\begin{table*}
  \renewcommand{\arraystretch}{1.2}

\caption{\label{tab_fraction_LAE} Table summarizing the interrelation between the LAE and LBG populations.
Numbers in boldface are absolute number of detections, and uncertainties correspond to the Poissonian error count.
}
\centering
\begin{tabular}{l|c|ccc}
\hline
 & $2.9\le z \le6.7$ & $2.9\le z \le 4.0$ &  4.0$\le z \le$5.0 &  5.0$\le z \le$6.7 \\
\hline
Effective Volume (Mpc$^3$)  & 13\,361 &  4\,546 & 3\,638 & 5\,177  \\ 
\hline
Sample 1: LBG with LAE & {\bf 92$\pm$9.6} & {\bf 43$\pm$6.6} & {\bf 27$\pm$5.2} & {\bf 22$\pm$4.7}   \\
                       & 16.9$\pm$1.7\%   & 13.2$\pm$2.0\%   & 18.0$\pm$3.5\% & 32.8$\pm$7.0\% \\
\hline
Sample 2: LBG only     & {\bf 406$\pm$20.1}     & {\bf 286$\pm$16.9 }& {\bf 105$\pm$10.2} &  {\bf 33$\pm$5.7}   \\
                       & 74.6$\pm$3.7\%   &  81.9$\pm$5.2\% & 70.0$\pm$6.8\% & 49.2$\pm$8.5\% \\
\hline
Sample 3: LAE only     & {\bf 46$\pm$6.8} & {\bf 16$\pm$4.0} & {\bf18$\pm$4.2} &{\bf 12$\pm$3.5} \\
                       & 8.4$\pm$1.2\%  & 4.9$\pm$1.2\%  & 12.0$\pm$2.8\%  & 17.9$\pm$5.2\% \\
\hline
Total                  & {\bf 544$\pm$23.3} & {\bf 327$\pm$18.1} & {\bf 150$\pm$12.2} & {\bf 67$\pm$8.2} \\
\hline
\end{tabular}
\end{table*}

\begin{figure}
   \centering
   \includegraphics[width=0.48\textwidth]{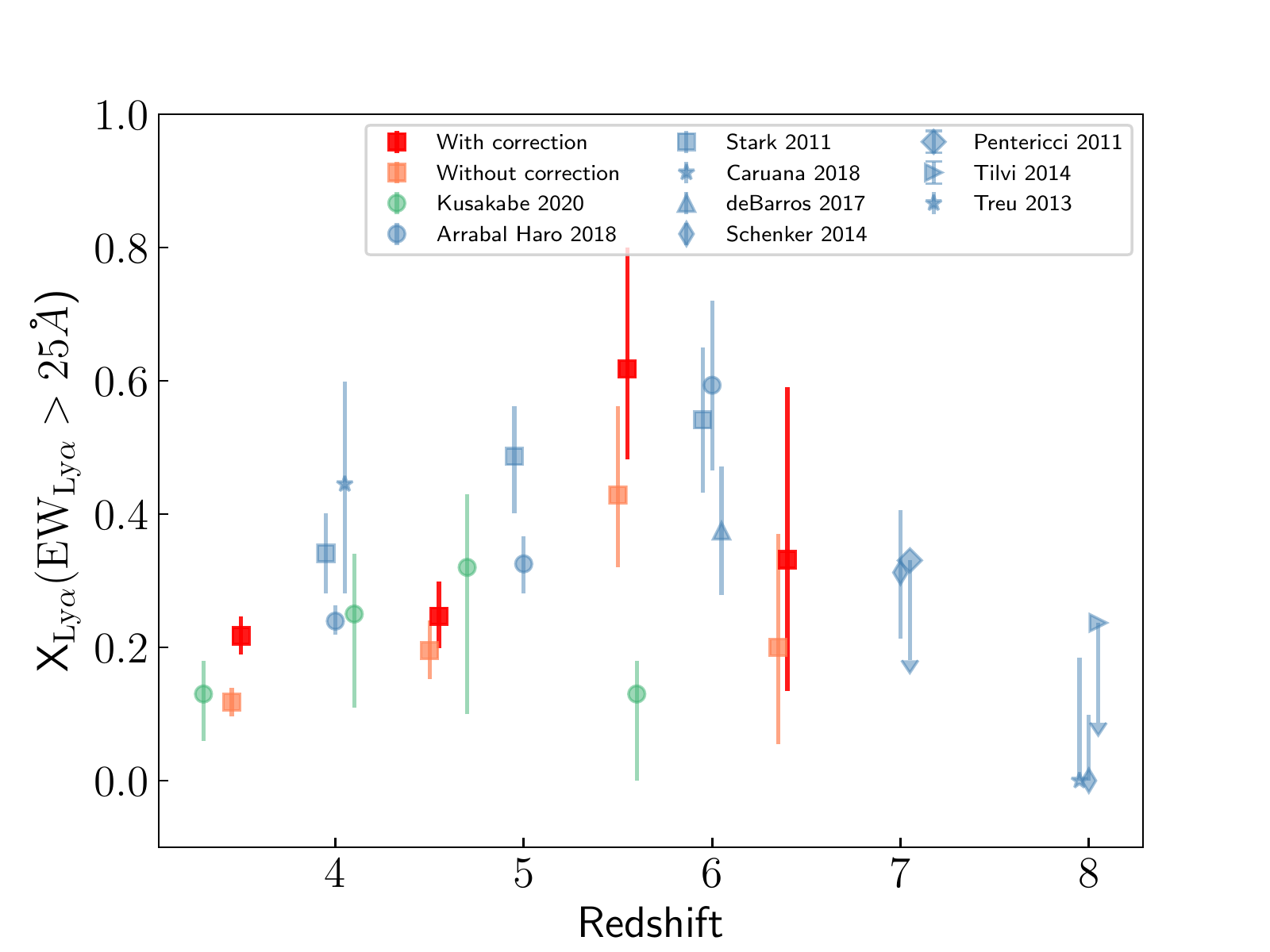}
   \caption{Fraction of LAEs with \ewlya $>$ 25 \AA \  
     among the UV selected galaxies with \Muv$>-20.25$. This fraction is computed from galaxies in Samples 1 and 2,
     and compared to other similar values in the literature (see Sect. \ref{evol_with_redshift} for more details). For clarity, a small redshift offset is applied to the points centered on the same redshift value. The definition of the \Muv range in the literature is not identical for the different authors.
It is $-20.25<$\Muv$<-18.75$ for \cite{Stark2011}, \cite{Pentericci2011} and  \cite{Kusakabe2020}, whereas it is \Muv$>-20.25$ in this work, as well as for \cite{Treu2013}, \cite{Schenker2014}, \cite{Tilvi2014}, \cite{deBarros2017}, \cite{Caruana2018} and \cite{ArrabalHaro2018}.
     }
     \label{fig_EW_fraction}
   \end{figure}

Using Sample 1 and 2 (LAE + LBG and LBG only) it is also possible to compute the fraction of LAEs  with \ewlya $>$ 25  \AA \ 
among our UV-selected sample, noted X$_{\rm LAE}$. This fraction, used to express the prevalence of LAEs  among the LBG population, is often divided into two populations: galaxies with \Muv $<-20.25$ and galaxies with \Muv$ > $-20.25 \citep[see e.g.][]{Stark2011,Pentericci2011,ArrabalHaro2018}. Because of the lensing nature of the present sample, it is mostly dominated by faint galaxies and 98\% of the sample falls within the \Muv$ > -20.25 $ domain. The limit \ewlya = 25\AA \ is shown in Fig.~\ref{fig_EW} and again most of our LAE sample falls above that limit. The LAE fraction is computed from Sample 1 and 2 with the cuts in both \ewlya and UV magnitude described above,
that is, using the upper limits for pure LBGs (Sample 2) at face values.
The results are presented with and without completeness correction for the LAE selection in Fig.~\ref{fig_EW_fraction} using the following redshift bins: $2.9<z<4.0$, $4.0<z<5.0$, $5.0<z<6.0$ and $6.0<z<6.7$. This binning was adopted to reach enough statistics in each bin and to easily compare with previous results, in particular around the $z\sim $ 6 epoch.

   The completeness of the LAE selection is determined using the method presented in \citet{deLaVieuville2019}. The individual maximum covolume of detection for each LAE is computed in the A2744 cube and is noted $V_{\max, i}$. This computation is done in the source plane, simulating the detectability of individual LAEs across numerous spectral layers (or monochromatic layers) and restricting the computation to the spatial areas where the magnification field is high enough to allow the detection of the LAE. An additional correction noted $C_i$ is considered to account for the fact that the LAE does not have a 100\% chance of being detected on its own spectral layer, due to the random variation of noise across the spatial dimension of the layer. This $C_i$ takes values in the range 0 - 1 and is computed on the detection layer of individual LAEs by injecting their own detection profile multiple times across the FoV and trying to recover it. All the technical details related to the computation of $V_{\rm max, i}$ and $C_i$ can be found in \citet{deLaVieuville2019}.

   The factor $1/C_iV_{\rm max, i}$ gives a numerical density for one LAE.
   To compute a correction from this numerical density, two additional quantities are introduced: the limit magnification noted $\mu_{\rm lim, i}$ and $V(>\mu)$. The limit magnification is the magnification value under which the S/N of a given LAE drops under one, and represents the minimum value of magnification for a specific LAE to be detected. The second one is the volume of observation explored above a given magnification, computed from the source plane. For each LAE, its corrected contribution to X$_{\rm LAE}$ is given by:

   \begin{equation}
     \label{eq:n_lae_corr_definition}
     N(\text{LAE})_{\rm corr, i } = \frac{V(>\mu_{\rm lim, i})}{C_i V_{\rm max, i}}.
   \end{equation}

   The term $V( > \mu_{\rm lim, i})$ is the volume of observations above a given amplification (i.e. neglecting the effects of the spectral variations of noise across the layers of the cube) and is used to normalize the intrinsic numerical density of the LAE. The point of such a normalization is that the background volume explored varies strongly depending on the magnification regime considered as faint sources can only be detected within the highly magnified regions. The correction of all LAEs with $N(\text{LAE}_{\rm corr, i }) \ge > 10 $ is neglected as it is estimated as untrustworthy. For a given UV-selected population, assumed to be complete, the fraction X$_{\rm LAE}$ is computed for two cases, the corrected and the uncorrected population of LAEs. Only Poissonian uncertainties are considered for the error propagation affecting  X$_{\rm LAE}$.

   In Fig.~\ref{fig_EW_fraction} the present results are compared to the ones derived by previous authors for \Muv$>-20.25$, namely \cite{Stark2011, Pentericci2011,Treu2013,Schenker2014,Tilvi2014,deBarros2017,ArrabalHaro2018}, 
keeping in mind that the definition of the \Muv range in the literature is not identical for the different authors. It is $-20.25<$\Muv$<-18.75$ for \cite{Stark2011} and \cite{Pentericci2011}, whereas it is \Muv$>-20.25$ in this work, as well as for \cite{Treu2013}, \cite{Schenker2014}, \cite{Tilvi2014}, \cite{deBarros2017}, \cite{Caruana2018} and \cite{ArrabalHaro2018}.
As expected, the present determination of X$_{\rm LAE}$ increases from $z = 3.5 $ to  $z = 5.5$ and drops for $z>$6, which can be interpreted
as an increase in the neutral fraction of hydrogen. Our corrected points are most consistent with the determination of X$_{\rm LAE}$ from \citet{ArrabalHaro2018}. Even though all of our corrected points are roughly consistent to 1$\sigma$ level with the other values from the literature, it appears that the two lower redshift points tend to be below previous estimates. Several factors may explain this observed trend:

\begin{itemize}
  
\item Because of the lensing nature of the present sample, the volume probed is small, only $\sim 13\,360$ Mpc$^3$ are explored in the range $2.9 < z < 6.7$, and the cosmic variance has therefore a high impact on the observed statistics. This additional uncertainty is not shown in the error bars.
  
\item There is clearly a difference in the selection processes with respect to previous studies. Here we use both broad-band photometry and IFU observations. The combination of these two methods ensures that we are as unbiased as possible in both the LBG and LAE selection in the same volume, range of magnitude and \Lya luminosity. All LAEs with a detected continuum are included in Sample 1, even if this detected continuum has a lower S/N that would not have allowed it to pass the photometric selection required for the pure LBGs of Sample 2 (this effect only accounts for $\sim$10$\%$ of Sample 1).
On one hand, two-step surveys, based on LAEs spectroscopically identified among photometrically pre-selected LBG populations, are likely missing LAEs with the faintest continuum counterparts, or exhibiting extended and/or ex-centered \Lya emission. On the other hand, \Lya emission is also likely to affect broad-band photometry used in the LBG pre-selection as well as the UV detection limit, as discussed in \cite{Kusakabe2020}. 
On the contrary, \mbox{\cite{ArrabalHaro2018}} used the 25 medium bands of the SHARDS survey to select both the LBG and LAE populations. These observations have an average depth of \mbox{$m = $ 26.5 - 29} magnitude and an average spectral resolution of $R \sim 50$. They are therefore mostly sensitive to UV-brighter galaxies and higher \ewlya, but are able to probe a much larger area. In this regard, this present study is complementary to their findings.

\item Finally, we get $\sim 40 \%$ more objects using a photo-z selection compared to a more traditional colour-colour selection (see beginning of Sect. \ref{selection}). This selection effect tends to lower our measurement of X$_{\rm LAE}$ compared to previous studies, but ensures that we have a more inclusive assessment of the high redshift SFGs.
  
\end{itemize}

 \subsection{Evolution with UV magnitude and \Lya luminosity}

 \begin{figure*}
   \centering
   \includegraphics[width=\textwidth]{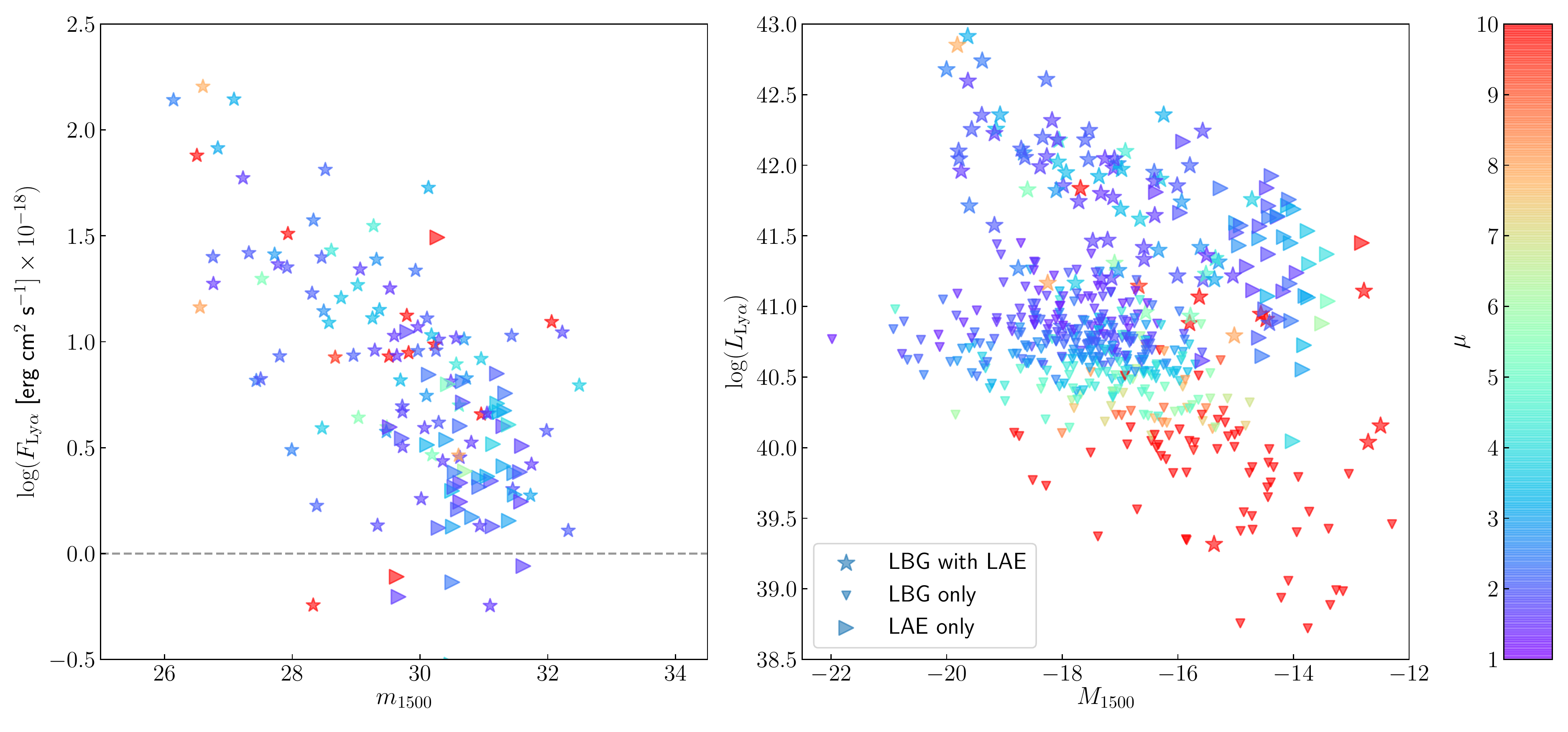}
   \caption{Distribution of the different populations of SFG in their observed and derived properties, showing the magnification $\mu$ affecting each galaxy in a color coded value (see color scale on the right).
The left panel displays the observed \Lya flux versus apparent magnitude $m_{1500}$, without any correction for magnification, for Samples 1 and 3. The $m_{1500}$ value has been recomputed based on the SED of each galaxy in Sample 1; for Sample 3, it is the limiting magnitude for each source as described in Sect. \ref{results}. The limiting \Lya flux assigned to all galaxies in Sample 2 is shown by a grey dashed line. 
The right panel displays the Lyman $\alpha$ luminosity versus absolute UV magnitude $M_{1500}$, after correction for magnification $\mu$. No correction for dust extinction was applied. The effect of magnification in spreading the distribution of galaxies towards the faintest values is clearly seen in this figure.}
         \label{fig_color_magnification}
\end{figure*}
 
 \begin{figure*}
   \centering
   \includegraphics[width=\textwidth]{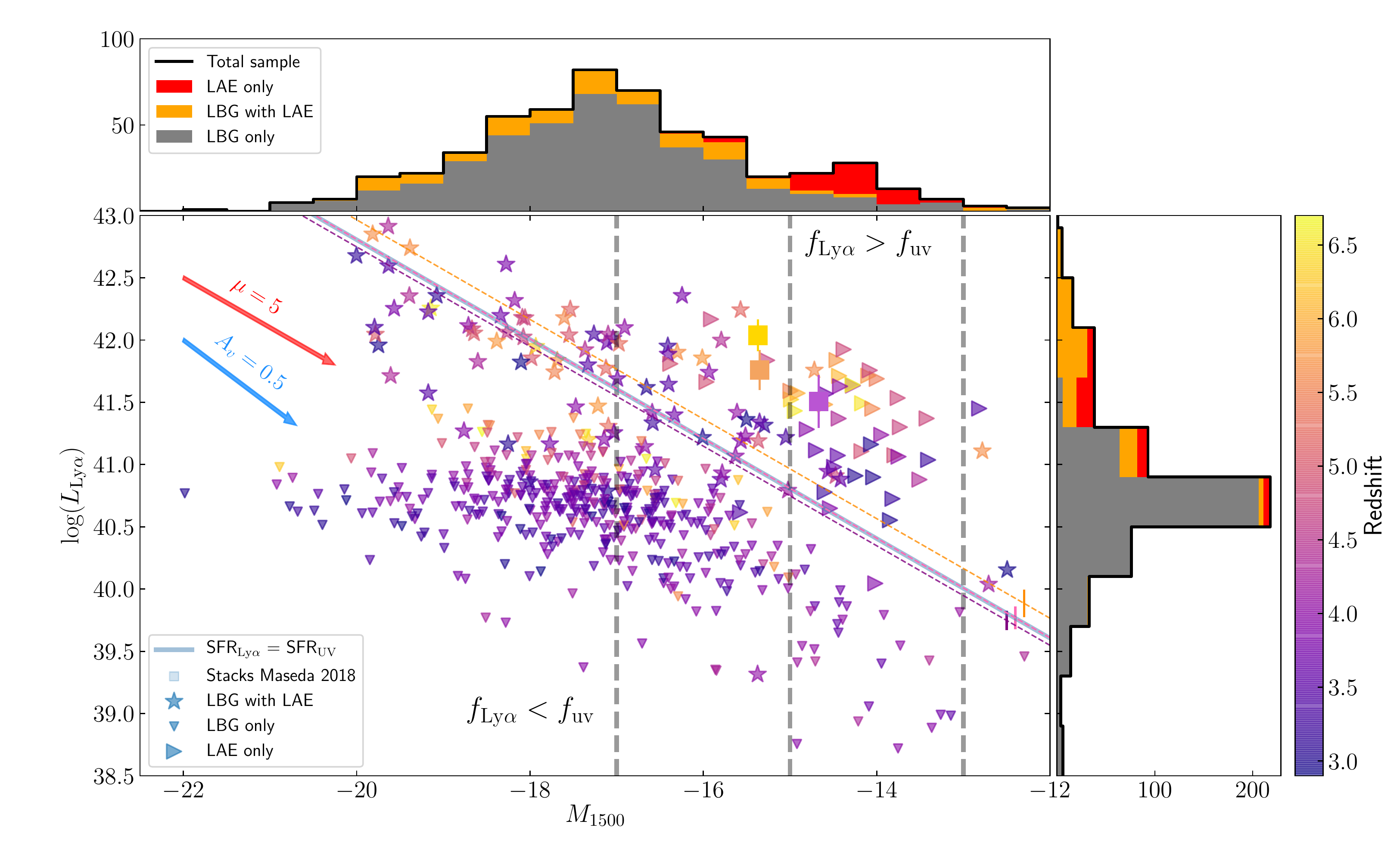}
   \caption{Lyman $\alpha$ luminosity versus absolute UV magnitude, with redshift encoded in the color bar. On the top (right) are UV absolute magnitude (\Lya luminosity) histograms for the three samples, stacked on each other, using the same colors as in Fig.~\ref{fig_histoz}. Galaxies in Samples 2 and 3 have upper limit estimations of their \Lya and UV fluxes (see text for details). All values have been corrected for magnification but not for possible dust extinction. The thick blue line displays the locus of the \sfrlya = \sfruv for a constant star-formation system. The dashed violet, pink and orange lines correspond to the best-fit for the offset in this relation (with fixed slope) when the sample is split in the redshift bins: $2.9 < z < 4.0$, $4.0< z <5.0$ and $5.0 < z <6.7$ respectively. Uncertainties on offset values are shown as vertical lines of the same color. The three vertical grey dashed lines correspond to \Muv$=-17$ \Muv$=-15$ \Muv$=-13$, which correspond roughly to the observational limit reached in blank fields, lensing fields, and the extrapolation needed to ensure that the UV LF provides enough ionizing flux for reionization \citep[see][]{Bouwens2015b}. Red and blue arrows indicate the effect of a magnification factor of $\mu=5$ and a reddening A$_{V}=0.5$ magnitudes respectively.}
         \label{fig_Lya_M1500}
\end{figure*}

 The corresponding \Lya luminosities and absolute UV magnitudes \Muv have been computed for all SFG in the three samples, including the magnification correction. \Lya luminosities are based on the observed \Lya fluxes and spectroscopic redshifts for Samples 1 and 3, and the limiting flux and photometric redshifts for Sample 2. Absolute magnitudes \Muv are based on the observed SED fit and spectroscopic (Sample 1) or photometric (Sample 2) redshifts, and limiting magnitudes and spectroscopic redshifts for Sample 3. 
   Figure.~\ref{fig_color_magnification} displays the distribution of the different populations of SFG in their observed and derived properties. This figure illustrates, on one hand, the effects of the magnification $\mu$ affecting each galaxy and, on the other hand, the impact of the limiting \Lya flux and $m_{1500}$ values assigned to Samples 2 and 3 respectively. On the left panel, the $m_{1500}$ value has been recomputed based on the SED of each galaxy in Sample 1; for Sample 3, it is the limiting magnitude for each source as described above. The effect of magnification in spreading the distribution of galaxies towards the faintest luminosities is clearly seen in the right panel, showing the Lyman $\alpha$ luminosity versus absolute UV magnitude $M_{1500}$, after correction for magnification $\mu$. As it is seen in this figure, given the limiting \Lya flux assigned to Sample 2, this population of pure LBGs tend to merge with the LBG$+$LAE population (Sample 1) for magnitudes \Muv$>-17$ to $-16$, the precise value depending on the somewhat arbitrary flux limit adopted for the non-detection (see beginning of Sect.~\ref{results}). The bulk of galaxies in Sample 3 is distributed between $-16<$\Muv$<-13$.

The main results of this study are summarized in Fig.~\ref{fig_Lya_M1500} which shows in its central part the three samples in a plot presenting the \Lya luminosity in log-scale versus \Muv, in a similar way as in Fig. \ref{fig_color_magnification}, with redshift color coded. UV magnitude and \Lya luminosity histograms of the three samples are also provided on the side.

   Regarding the star formation rate (SFR), Fig.~\ref{fig_Lya_M1500} also displays the locus of the
   \sfrlya $ = $ \sfruv for a constant star-formation system.
This relation was computed using the standard conversion in \cite{Kennicutt1998}
for \sfrlya, and the common conversion also given in \cite{Kennicutt1998} based on a
Salpeter IMF for the \sfruv. Since none of the values presented in Fig.~\ref{fig_Lya_M1500} have been corrected for dust absorption, this constant star-formation line also represents the locus where the escape
fraction for \Lya photons is the same as for UV photons (\flya = \fuv). Regarding the total ionizing flux density (or the Star Formation Rate density, SFRD), it means that objects along this line yield the same values when measuring either the UV continuum or the \Lya line flux. Galaxies found above or below this line
in this simple model have respectively \fuv < \flya or \fuv > \flya. In addition, the scatter can be easily understood given the variety of star-forming regimes that are not fairly represented by a stationary and constant star-formation regimes. 

The three samples are each exploring different regions of Fig. \ref{fig_Lya_M1500}. For bright galaxies with \Muv$<-15$,
almost all SFGs in the present sample are selected as LBGs.
However, for galaxies fainter than \Muv$=-15$ the proportion of SFGs only seen as LAEs progressively increases at lower UV luminosity. This result is in good agreement with the recent study of \cite{Maseda2018} which concluded that the LAE selection is better suited to identify and study intrinsically faint UV galaxies.  This also suggests that the LAE Luminosity Function is a better proxy of the SFG population when focusing on the faint end. The observed trend depends on the relative depth between MUSE and \textit{HST} observations: deeper \textit{HST} observations would increase the number of faint LBGs detected. Needless to say that such observations are extremely expensive and the HFF data are currently the deepest observations available.

In addition to the theoretical \sfrlya = \sfruv locus, the same relation has been adjusted with the offset as a free parameter, keeping a constant slope, for galaxies in Sample 1 split in the following redshift bins: $2.9< z <4.0$ (43 sources), $4.0 < z < 5.0$ (27 sources) and $5.0 < z < 6.7$ (22 sources). The results are presented respectively by the dashed violet, pink and orange lines in Fig.~\ref{fig_Lya_M1500}, where the derived uncertainties from the fit are represented as vertical error bars of the same color. Leaving this offset free means leaving the ratio \rf free, in such a way that the fit can be seen as an average value of this ratio over the sample considered. For the two lower redshift bins, the adjusted line is fully consistent with the \sfrlya = \sfruv locus within the 1$\sigma$ level, whereas a \flya > \fuv is observed for the higher redshift bin with a $\sim 1\sigma$ significance.

When looking at galaxies of Sample 1, on average, ``bright'' galaxies with \Muv$<-17$ tend to be under the \sfrlya = \sfruv locus, therefore extending a trend already observed for brighter samples at such high-redshifts
\citep[see e.g.][]{Ando2006,Schaerer2011}. 
Given the limiting detection levels in this survey, almost all galaxies in Sample 3 are located above the \sfrlya = \sfruv locus, as well as the faintest galaxies of Sample 1 with \Muv$>-17$, in such a way that the number of galaxies with \sfrlya $>$ \sfruv at \Muv$>-17$ is larger than for \Muv$<-17$. 
Such a behavior could be explained by the evolution of the dust content, and the relative distribution of dust and stars in star-forming regions, as discussed below. Also the large scatter observed
in Sample 1 for all the three redshift bins could be representative of individual variation on the \rf ratio related to the star formation regime, the age of the stellar population, the dust content, and the relative distribution of stars and dust.\\

   It is worth emphasizing that the LBG sample used throughout this work is NIR-selected (i.e. selected based on the rest frame continuum at $\sim$4000\AA \
   down to 2000\AA \ 
   for $z\sim$3 to 7 respectively). We expect a systematic trend in the sense that LBG galaxies with extremely blue continuum could have been preferentially missed at $z \sim$ 3-4 with respect to $z \sim$ 6-7. While this certainly impacts our statistics, this effect can hardly account for the systematic trends presented above.

   The fact that all SFGs with \Muv$<-17$ in the present sample are selected as LBGs is particularly interesting since it roughly corresponds to the observational limit of the deepest \textit{HST} blank fields \citep[see e.g.][]{Bouwens2015b}.
This limit can be pushed as faint as \Muv$\sim-15$ in strong lensing fields
before a significant correction is required to account for the contribution of LAEs without continuum detection.  This means that the UV LF derived from LBG selection can be safely integrated down to \Muv$\sim-15$ to derive the total ionizing flux without including the additional contribution of the (pure) LAE population.
Reaching down to \Muv$\sim-15$ up to $z \sim7 $ is only possible in lensing fields, as shown by different studies \citep[see e.g.][]{Livermore2017,Bouwens2017,Atek2018}.
The situation is different for \Muv$>-15$, as the pure LAE fraction among SFGs increases towards the faintest UV luminosities. Since the \Muv values and associated luminosities for pure LAE galaxies are upper limits, it is not possible to determine individually whether they probe the same luminosity domain as (extremely faint) LBGs
not detected as LAEs (Sample 2). Therefore, it is difficult to properly estimate the relative contribution to the ionizing flux for the ensemble of SFGs.
For illustration purposes, we provide an upper limit on the fraction of LAEs among all SFGs in the range $-15 \le$ \Muv $\le -14$ where the completeness of the UV selection starts to drop. Using the lower limits on \Muv derived for Sample 3, the computed upper limit and $1-\sigma$ values are $N(\text{LAE})/N(\text{SFG}) = 0.42 \pm 0.14 $.

Stacking the galaxies of Sample 3 in redshift bins,
following the technique adopted by \cite{Maseda2018}
would allow us to know more on their average UV continuum. But this remains challenging because the strong variation of individual amplification makes it non trivial to give similar weight to all stacked LAEs; also the sample of pure LAEs is still very small. Therefore, based on the present results it is still difficult to quantify the missing contribution of LAE to the ionizing flux density with respect to the extrapolation of the LBG-based UV LF to
\Muv$\sim - 13$, if any, up to $z\sim 7$.

For the sample of SFGs studied here, bright galaxies with \Muv$<-17$ tend to be under the \sfrlya = \sfruv locus, whereas galaxies with \Muv$>-17$ are mostly observed above. Also a higher \sfrlya (with respect to \sfruv) is observed for galaxies at $z >$5 as compared to $z <$5 but, as discussed above, this effect is only at $\sim 1\sigma$ significance for this sample. Several effects, combined together, could explain these observations in the main trends and the scatter, including a genuine evolution in the \rf ratio:

  \begin{itemize}
    
  \item Possibly the simplest explanation for these trends is the age of the stellar populations. As previously suggested by \cite{Ando2006}, if UV ``bright'' galaxies start their star-formation earlier than UV ``faint'' galaxies, they are expected to contain older stellar populations and be more chemically evolved and dustier, which could explain a relative decrease in their \Lya emission. Conversely, for \Muv$>-17$ galaxies, the presence of a younger stellar population producing a harder spectrum, with lower metallicity stars, is expected to result in a the relative increase of the \Lya emission. This effect was also discussed in the past \citep[see e.g.][]{Erb2014}. The original result here is that these effects are seen for very faint \Muv$\sim -15$ galaxies, much fainter than in previous studies.
Regarding the scatter, even in the case of a constant star formation rate, the ratio of UV to \Lya luminosity produced by a young stellar populations evolves and needs $\sim 100$ Myr before reaching the Kennicut equilibrium \citep{Verhamme2008}. The same is true in the case of a more realistic non constant and bursty SFR, which could explain part of the observed scatter.
    
  \item Regarding the relative increase of \sfrlya with respect to \sfruv for the faintest galaxies (\Muv$>-17$), an interesting explanation is the scenario described by \cite{Neufeld1991}, also suggested by \cite{Atek2014}, based on the existence of a multi-phase ISM with dust, where the \Lya emission is enhanced with respect to the non-resonant emission at similar wavelengths (i.e. UV continuum photons in the present case). Following this scenario, neutral gas and dust reside in clumps and they are surrounded by an ionized medium. The \Lya photons are scattered away when reaching the surface of these clumps  while the UV photons can penetrate inside where they are more easily absorbed by the dust. Therefore even though \flya decreases with increasing dust content or reddening, the ratio \rf could increase since \fuv decreases faster than \flya in presence of dust clumps. Because, as predicted by models \citep[see e.g.][]{Garel2012}, faint UV galaxies tend to be less dusty and more clumpy than ``bright'' galaxies, this effect is expected to preferentially impact the faintest population. Several previous studies have mentioned the importance of this mechanism \citep[see e.g.][]{Hayes2011,Hayes2013,Atek2014,Matthee2016}. However, as pointed out by \cite{Duval2014} based on their 3D Monte Carlo \Lya radiation transfer code, special conditions could be needed to have \Lya photons escaping more easily than the continuum (i.e. an almost static ISM, extremely clumpy and very dusty). We do not have enough elements to know the real impact of this effect on our sample. 

  \item Geometrical effects are also important and certainly contribute to the observed scatter. A more uniform dust distribution makes the absorption of \Lya photons more likely leading to \fuv > \flya, a systematic trend observed in the present sample for \Muv$<-17$ galaxies. Simulations by \cite{Verhamme2012}, studying the effect of disk orientation, have shown the dramatic effect of inclination on the \Lya observed properties (luminosity, EW and profile), with luminosity increasing from edge-on to face-on due to a smaller number of the resonant scattering, these effects being much stronger for \Lya photons than for the UV continuum. Observations by \cite{Shibuya2014} of LAEs at z$\sim$2.2 confirm this inclination trend.  
Several authors have pointed out the fact that the neutral hydrogen column density is a really key value affecting the \Lya emissivity \citep[see e.g.][]{Shibuya2014b,Hashimoto2015}. For galaxies with the lower UV luminosities (likely the smaller masses), the observed relative excess in the \flya could be related to the fact that large gaseous disks are not yet in place, in addition to possible age effects, as mentioned above \citep[see e.g.][]{Erb2014}. 

\item Additional differential effects on the \Lya emissivity are expected from different physical conditions affecting the nebular regions, in particular the ionization states, if these regions are density rather than ionization-bounded. Such effects have been described and considered by several authors when discussing the spectral signatures expected and observed for SFGs at high-$z$ \citep[see e.g.][]{Zackrisson2013,Nakajima2016,Shibuya2019}.
  
\end{itemize}



\section{Conclusions}
\label{conclusion}

We have studied the intersection between the LAE and the LBG population behind the HFF cluster A2744. This has resulted in the following conclusions:
\begin{itemize}
\item For faint UV-selected galaxies with \Muv$\ge -20.25$, the fraction
of LAE among SFGs increases with redshift up to $z \sim$6 and decreases at $z \gtapprox 6$ ,
in agreement with previous findings \citep[see e.g.][]{ArrabalHaro2018}.

\item As faint as \Muv$\sim -15$, the LBG population seems to provide a good representation of the total SFG population, in particular when computing the total ionizing flux in the volume explored by current surveys.

\item The selection of \Lya emitters allows us to detect a population of intrinsically UV faint galaxies (\Muv$\ge-15$) with significant star-formation (typically 0.01 to 0.1 M$_\odot$/yr) that are missed in present deep blank and lensing field LBG photometric surveys.
In this respect, our results are in good agreement with \cite{Maseda2018}. This also shows that when assessing the faint part of the population of SFG with \Muv$\ge-15$ with the current deepest photometric data, a correction is needed to account for the contribution of the LAEs with no UV counterpart detection.

\item There is no clear evidence, based on the present results, for an intrinsic difference on the properties of
the two populations selected as LBG and/or LAE. However, further investigation will be needed. In particular, some systematic trends appear in the population selected as LBG and LAE, in the sense that the UV-brightest galaxies seem to exhibit a smaller \rf ratio, increasing towards the faintest luminosities. 
These trends, previously observed in other samples of SFGs at high-$z$ 
\citep[see e.g.][]{Ando2006,Erb2014,Hashimoto2017},
are seen here for very faint \Muv$\sim -15$ galaxies, much fainter than in previous studies.
Several effects acting in combination could explain these differential trends, such as the age of the stellar populations, a different distribution of ISM and stars, geometrical effects and physical conditions affecting the nebular regions. 

Measuring
the UV-slopes of these galaxies could provide additional information in this respect.
\end{itemize}

\begin{acknowledgements}
Partially funded by the ERC starting grant CALENDS (JR),
the Agence Nationale de la recherche bearing the
reference ANR-13-BS05-0010-02 (FOGHAR), and the
``Programme National de Cosmologie and Galaxies" (PNCG) of CNRS/INSU
with INP and IN2P3, co-funded by CEA and CNES, France.
This work has been carried out thanks to the support of the OCEVU Labex
(ANR-11-LABX-0060) and the A*MIDEX project (ANR-11-IDEX-0001-02) funded by the
"Investissements d'Avenir" French government program managed by the ANR.
This work also recieved support from the  ECOS SUD Program C16U02.
NL acknowledges support from the Kavli foundation. FEB acknowledges support from CONICYT grants CATA-Basal AFB-170002 (FEB), 
FONDECYT Regular 1190818 (FEB), and Programa de Cooperaci{\'{o}}n Cient{\'{\i}}fica ECOS-CONICYT C16U02 the Chilean Ministry of Economy, Development, and Tourism's Millennium Science Initiative through grant IC120009, awarded to The Millennium Institute of Astrophysics, MAS. TG acknowledges support from the European Research Council under grant agreement ERC-stg-757258 (TRIPLE). Based on observations made with ESO Telescopes at the La Silla Paranal Observatory under programme ID 094.A-0115.
\end{acknowledgements}

\bibliographystyle{aa} 
\bibliography{bibliography_copy} 

\end{document}